\documentclass[12pt]{article}
\usepackage{epsf} 
\usepackage{epsfig} 
\usepackage{wrapfig}

\catcode`\@=11
\long\def\@makefntext#1{
\protect\noindent \hbox to 3.2pt {\hskip-.9pt  
$^{{\ninerm\@thefnmark}}$\hfil}#1\hfill}                

\def\@makefnmark{\hbox to 0pt{$^{\@thefnmark}$\hss}}  
        
\def\ps@myheadings{\let\@mkboth\@gobbletwo
\def\@oddhead{\hbox{}
\rightmark\hfil\ninerm\thepage}   
\def\@oddfoot{}\def\@evenhead{\ninerm\thepage\hfil
\leftmark\hbox{}}\def\@evenfoot{}
\def\sectionmark##1{}\def\subsectionmark##1{}}

\renewcommand{\thefootnote}{\fnsymbol{footnote}}

\newcounter{sectionc}\newcounter{subsectionc}\newcounter{subsubsectionc}
\renewcommand{\section}[1] {\vspace*{0.6cm}\addtocounter{sectionc}{1} 
\setcounter{subsectionc}{0}\setcounter{subsubsectionc}{0}\noindent 
        {\normalsize\bf\thesectionc. #1}\par\vspace*{0.4cm}}
\renewcommand{\subsection}[1] {\vspace*{0.6cm}\addtocounter{subsectionc}{1} 
        \setcounter{subsubsectionc}{0}\noindent 
        {\normalsize\it\thesectionc.\thesubsectionc. #1}\par\vspace*{0.4cm}}
\renewcommand{\subsubsection}[1]
{\vspace*{0.6cm}\addtocounter{subsubsectionc}{1}
        \noindent {\normalsize\rm\thesectionc.\thesubsectionc.\thesubsubsectionc. 
        #1}\par\vspace*{0.4cm}}

\newcounter{appendixc}
\newcounter{subappendixc}[appendixc]
\newcounter{subsubappendixc}[subappendixc]

\renewcommand{\appendix}[1] {\vspace*{0.6cm}
        \refstepcounter{appendixc}
        \setcounter{figure}{0}
        \setcounter{table}{0}
        \setcounter{equation}{0}
        \renewcommand{\thefigure}{\Alph{appendixc}.\arabic{figure}}
        \renewcommand{\thetable}{\Alph{appendixc}.\arabic{table}}
        \renewcommand{\theappendixc}{\Alph{appendixc}}
        \renewcommand{\theequation}{\Alph{appendixc}.\arabic{equation}}
        \noindent{\bf Appendix \theappendixc #1}\par\vspace*{0.4cm}}

\def\abstracts#1{{
        \centering{\begin{minipage}{12.2truecm}\footnotesize\baselineskip=12pt\noindent
        \centerline{\footnotesize ABSTRACT}\vspace*{0.3cm}
        \parindent=0pt #1
        \end{minipage}}\par}} 

\renewenvironment{thebibliography}[1]
        {\begin{list}{\arabic{enumi}.}
        {\usecounter{enumi}\setlength{\parsep}{0pt}
\setlength{\leftmargin 1.25cm}{\rightmargin 0pt}
         \setlength{\itemsep}{0pt} \settowidth
        {\labelwidth}{#1.}\sloppy}}{\end{list}}

\newcounter{itemlistc}
\newcounter{romanlistc}
\newcounter{alphlistc}
\newcounter{arabiclistc}

\newcommand{\fcaption}[1]{
        \refstepcounter{figure}
        \setbox\@tempboxa = \hbox{\footnotesize Fig.~\thefigure. #1}
        \ifdim \wd\@tempboxa > 6in
           {\begin{center}
        \parbox{6in}{\footnotesize\baselineskip=12pt Fig.~\thefigure. #1}
            \end{center}}
        \else
             {\begin{center}
             {\footnotesize Fig.~\thefigure. #1}
              \end{center}}
        \fi}

\newcommand{\tcaption}[1]{
        \refstepcounter{table}
        \setbox\@tempboxa = \hbox{\footnotesize Table~\thetable. #1}
        \ifdim \wd\@tempboxa > 6in
           {\begin{center}
        \parbox{6in}{\footnotesize\baselineskip=12pt Table~\thetable. #1}
            \end{center}}
        \else
             {\begin{center}
             {\footnotesize Table~\thetable. #1}
              \end{center}}
        \fi}

\def\@citex[#1]#2{\if@filesw\immediate\write\@auxout
        {\string\citation{#2}}\fi
\def\@citea{}\@cite{\@for\@citeb:=#2\do
        {\@citea\def\@citea{,}\@ifundefined
        {b@\@citeb}{{\bf ?}\@warning
        {Citation `\@citeb' on page \thepage \space undefined}}
        {\csname b@\@citeb\endcsname}}}{#1}}

\newif\if@cghi
\def\cite{\@cghitrue\@ifnextchar [{\@tempswatrue
        \@citex}{\@tempswafalse\@citex[]}}
\def\citelow{\@cghifalse\@ifnextchar [{\@tempswatrue
        \@citex}{\@tempswafalse\@citex[]}}
\def\@cite#1#2{{$\null^{#1}$\if@tempswa\typeout
        {IJCGA warning: optional citation argument 
        ignored: `#2'} \fi}}

 1
 1
 1

\font\ninerm=cmr9

\def\ee{\mbox{e}^+\mbox{e}^-}

\def\sqee{\sqrt{s}_{\rm ee}}

\def\gg{\gamma\gamma}

\def\ee{\mbox{e}^+\mbox{e}^-}

\newcommand{\sleq} {\raisebox{-.6ex}{${\textstyle\stackrel{<}{\sim}}$}}

\def\ETJET{E^{\rm jet}_T}

\def\qqbar{\mbox{q}\overline{\mbox{q}}}
\def\ccbar{\mbox{c}\overline{\mbox{c}}}

\def\xgp{x_{\gamma}^+}
\def\xgm{x_{\gamma}^-}
\def\xgpm{x_{\gamma}^{\pm}}
\def\etajet{\eta^{\rm jet}}

\def\Zzero{\ifmmode {{\mathrm Z}^0} \else {${\mathrm Z}^0$} \fi}
\def\ppbar{\overline{\mbox p}\mbox{p}}

\def\fl         {\ifmmode {F^{\gamma}_L}
                 \else    {$F^{\gamma}_L$}\fi}
\def\fb         {\ifmmode {F^{\gamma}_B}
                 \else   {$F^{\gamma}_B$}\fi}
\def\sigmagg{\sigma_{\gg}}
\begin{document}
\pagestyle{myheadings}
\markboth{FREIBURG-EHEP-97-15}{FREIBURG-EHEP-97-15}

\centerline{\normalsize\bf PHOTON STRUCTURE AT LEP\footnotemark[1] }
\baselineskip=15pt

\centerline{\footnotesize STEFAN S\"OLDNER-REMBOLD}
\baselineskip=13pt
\centerline{\footnotesize\it Universit\"at Freiburg}
\baselineskip=12pt
\centerline{\footnotesize\it 
Hermann-Herder-Str.~3, D-79104 Freiburg i.~Br., Germany}
\centerline{\footnotesize E-mail: soldner@ruhpb.physik.uni-freiburg.de}

\vspace*{0.5cm}
\abstracts{The structure of the photon is studied in photon interactions
at high energies using photons radiated by the electron and positron 
beams at LEP. The current status of these measurements is reviewed.
}
 
\normalsize\baselineskip=15pt
\setcounter{footnote}{0}
\renewcommand{\thefootnote}{\alph{footnote}}
\section{Introduction}
The photon is one of the fundamental gauge bosons of the Standard Model. 
At high energies photon interactions 
are dominated by quantum fluctuations
of the photons into fermion-antifermion pairs and into
vector mesons which have the same spin-parity ($J^{PC}=1^{--}$)
as the photon. This is called photon structure. 
Electron-Positron collisions at LEP are an ideal laboratory
for studying photon structure in interactions of quasi-real and virtual 
photons, testing predictions of both QED and QCD. 
These results are complementary to the results 
obtained in $\gamma$p scattering at HERA.

\section{Electron--Photon Scattering}
If one of the scattered electrons\footnote{In this paper positrons
are also referred to as electrons}~~is detected (tagged), the process
$\ee \rightarrow \ee + \mbox{hadrons} $ can be regarded
as deep-inelastic scattering of an electron on a quasi-real
photon with the cross-section
 \begin{equation}
  \frac{{\rm d}^2\sigma_{\rm e\gamma\rightarrow {\rm e+hadrons}}}{{\rm d}
x{\rm d}Q^2}
 =\frac{2\pi\alpha}{x\,Q^{4}}
  \left[ \left( 1+(1-y)^2\right) F_2^{\gamma}(x,Q^2) - y^{2}F_{\rm L}(x,Q^2)\right],
\label{eq-eq1}
 \end{equation}
where $Q^2=-q^2$ is the negative four-momentum squared of the
virtual photon and $x$ and $y$ are the usual dimensionless variables.
The structure function $F_2^{\gamma}$ is related to the sum over the parton 
densities of the photon. In order to identify an electron
in the detector, a large tag energy $E_{\rm tag}$ has to be required, 
i.e~only low values of $y$ are accessible ($y^2\ll 1$). 
The contribution of the term proportional to the longitudinal structure 
function $F_{\rm L}^{\gamma}$ is therefore negligible.
\setcounter{footnote}{0}
\renewcommand{\thefootnote}{\fnsymbol{footnote}}
\footnotetext[1]{Invited talk given at the Ringberg Workshop on New
Trends in HERA Physics, Ringberg Castle, Germany, 25-30 May 1997} 
 
For measuring $F_2^{\gamma}(x,Q^2)$ the value of $Q^2$ can
be well reconstructed from the energy, $E_{\rm tag}$,
and the angle, $\theta_{\rm tag}$, of the tagged electron via
the relation $Q^2\approx 2E_{\rm beam} E_{\rm tag}(1-\cos\theta_{\rm tag})$.
The reconstruction of $x=Q^2/(Q^2+W^2)$, however, relies heavily on the 
measurement of the invariant mass $W$ from the hadronic final state.

\subsection{Hadronic Energy Flows}
OPAL~\cite{bib-hadopal}, ALEPH~\cite{bib-alephflow} and 
DELPHI~\cite{bib-igor} have therefore studied 
the hadronic energy flow $1/N\cdot {\rm d}E/{\rm d}\eta$
as a function of the pseudorapidity $\eta=-\ln\tan\theta/2$, where
the sign of $\eta$ is chosen in such way that the
tag electron is always at negative $\eta$. OPAL has measured 
\begin{figure}[htbp]
\begin{tabular}{cc}
\epsfig{file=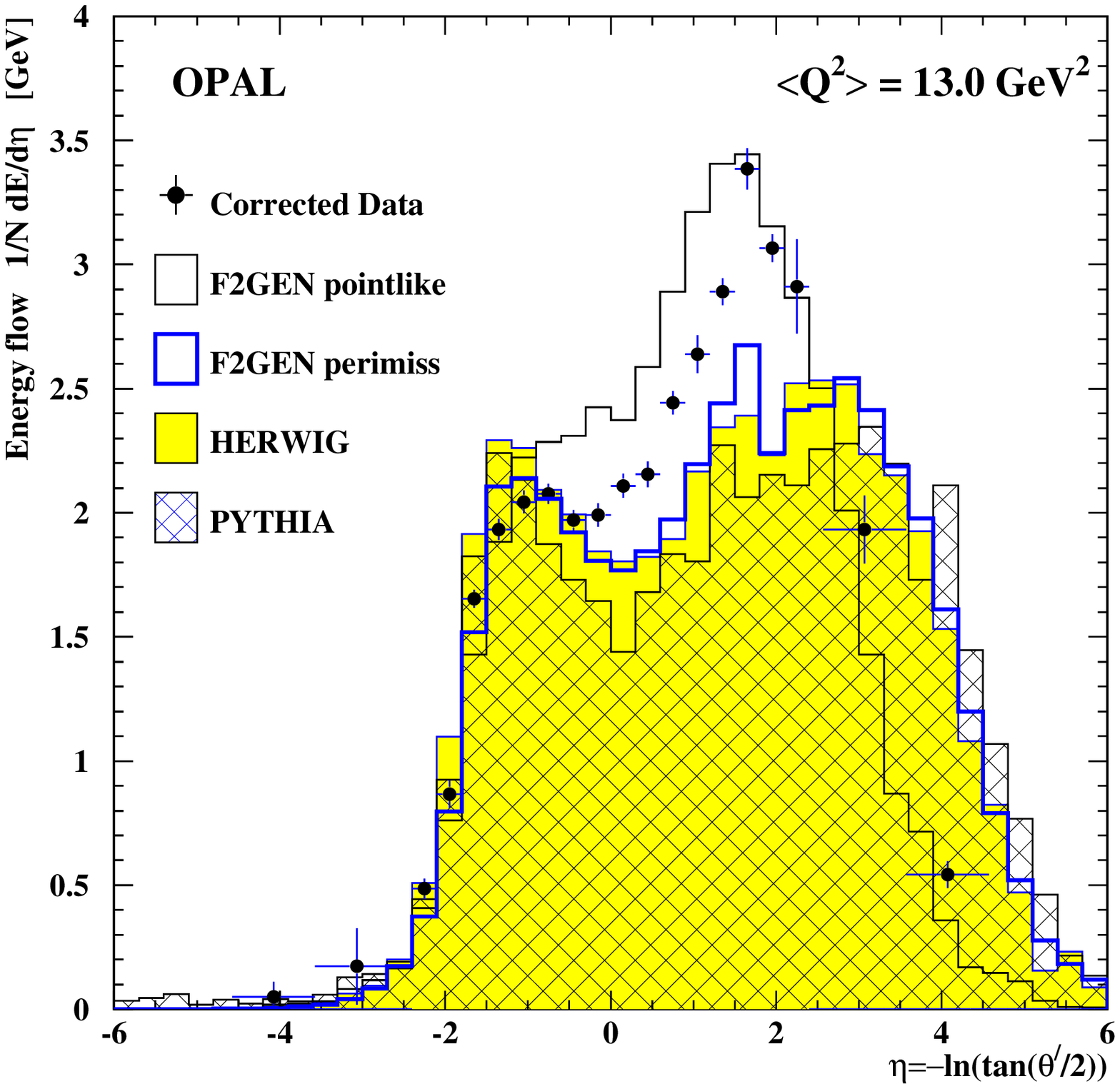,width=0.463\textwidth,height=5.3cm}
 &
\epsfig{file=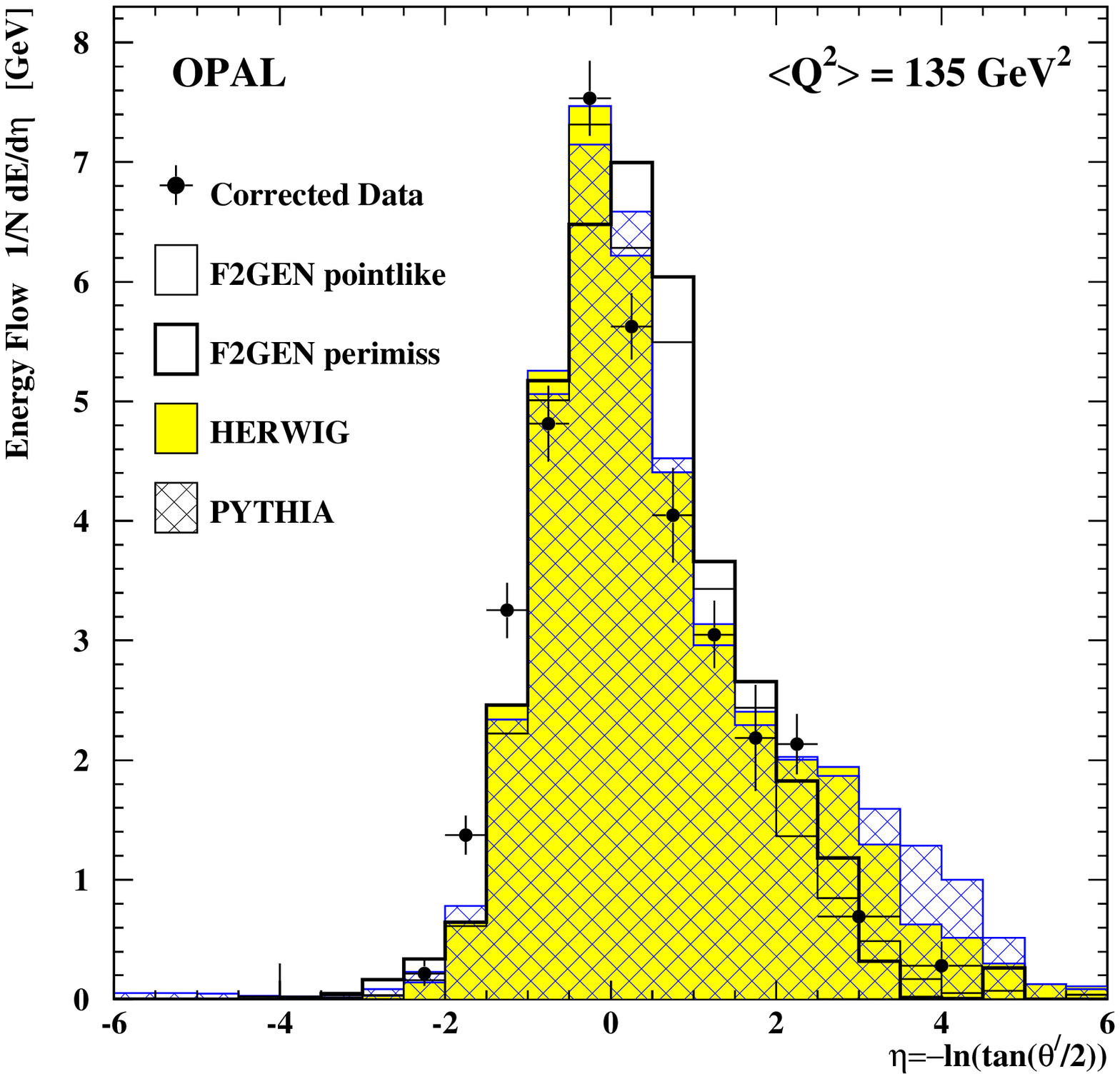,width=0.463\textwidth,height=5.3cm} 
\end{tabular}
\caption{\label{fig-hadopal}
 The energy flow per event
 as a function of the pseudorapidity $\eta$, compared to
 the HERWIG, PYTHIA and F2GEN Monte Carlo models.
 The data have been corrected for detector effects.}
\end{figure}
\begin{wrapfigure}{r}{2.5in}
\epsfig{file=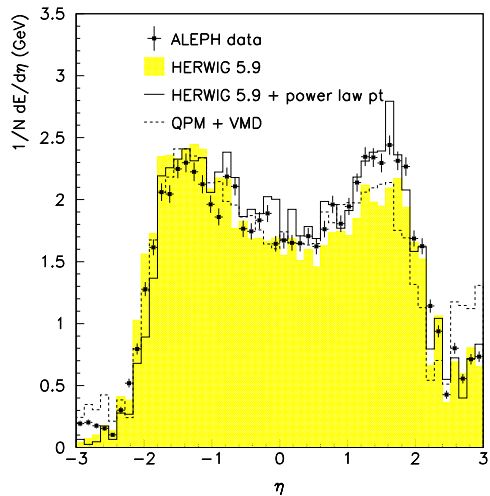,width=2.5in} 
\caption{\label{fig-alephflow}
 The energy flow per event
 as a function of $\eta$, compared to
 a QPM+VMD model,
 the standard and the tuned version of HERWIG.} 
\end{wrapfigure}
the energy flow at medium $Q^2$ 
($\langle Q^2\rangle=13$ GeV$^2$) and at high $Q^2$
($\langle Q^2\rangle=135$ GeV$^2$)~\cite{bib-hadopal}. 
In Fig.~\ref{fig-hadopal} the energy flows are compared
to the two QCD based Monte Carlo generators HERWIG~\cite{bib-herwig}
and PYTHIA~\cite{bib-pythia}.
The data distributions have been corrected for detector effects.
The generator F2GEN is used
to model the unphysically extreme case 
of a two-quark state in the $\gamma^*\gamma$ centre-of-mass system
with an angular distribution as in lepton pair production from two real 
photons (``pointlike''). The ``perimiss'' sample is 
a physics motivated mixture of pointlike and peripheral
interactions, where peripheral means that the
transverse momentum of the outgoing quarks is given by
an exponential distribution as if 
all the photons interacted as pre-existing hadrons.

Significant discrepancies exist between the data and all of
the Monte Carlo models. The agreement improves at higher $Q^2$.
Since $x$ and $Q^2$ are correlated, the discrepancies
at low $Q^2$ are observed also at low $x$. These discrepancies
between the data and the Monte Carlo model for the hadronic final
state are the dominant source of systematic uncertainty
in the unfolding of $F_2^{\gamma}(x,Q^2)$~\cite{bib-hadopal}.

Tuning of the Monte Carlo to improve the modelling of
the hadronic final state is complicated and must be done with care
in order to avoid a bias of the result of the unfolding
towards the parametrisation of the parton distribution
functions used in the tuned Monte Carlo. ALEPH has
measured the energy flow in tagged event for
$\langle Q^2 \rangle=14.2$~GeV$^2$~\cite{bib-alephflow}.
The distributions have not been corrected for detector effects.
The energy flow shown in Fig.~\ref{fig-alephflow} is compared
to the HERWIG generator~\cite{bib-herwig} and a Monte Carlo
which consists of a mixture of Quark Parton Model (QPM)
and Vector Meson Dominance (VMD) similar to the F2GEN
generator. In addition, a modified version
of HERWIG (``HERWIG+power law $p_{\rm T}$'') was used.
The modification is based on studies of the photon remnant by 
ZEUS~\cite{bib-rzeus}. In standard HERWIG a Gaussian distribution is 
used to describe
the limitation of the transverse momentum of the outgoing partons
with respect to the initial target photon. In the modified
HERWIG the Gaussian is replaced by a power law spectrum.
The agreement with the data improves. A similar study
was performed earlier by Lauber in Ref.~7 using OPAL data.
It is expected that such improvements of the Monte Carlo
models will significantly reduce the systematic error
of the structure function measurements.

\subsection{Photon Structure Function}
\begin{figure}[htbp]
\begin{tabular}{cc}
\epsfig{file=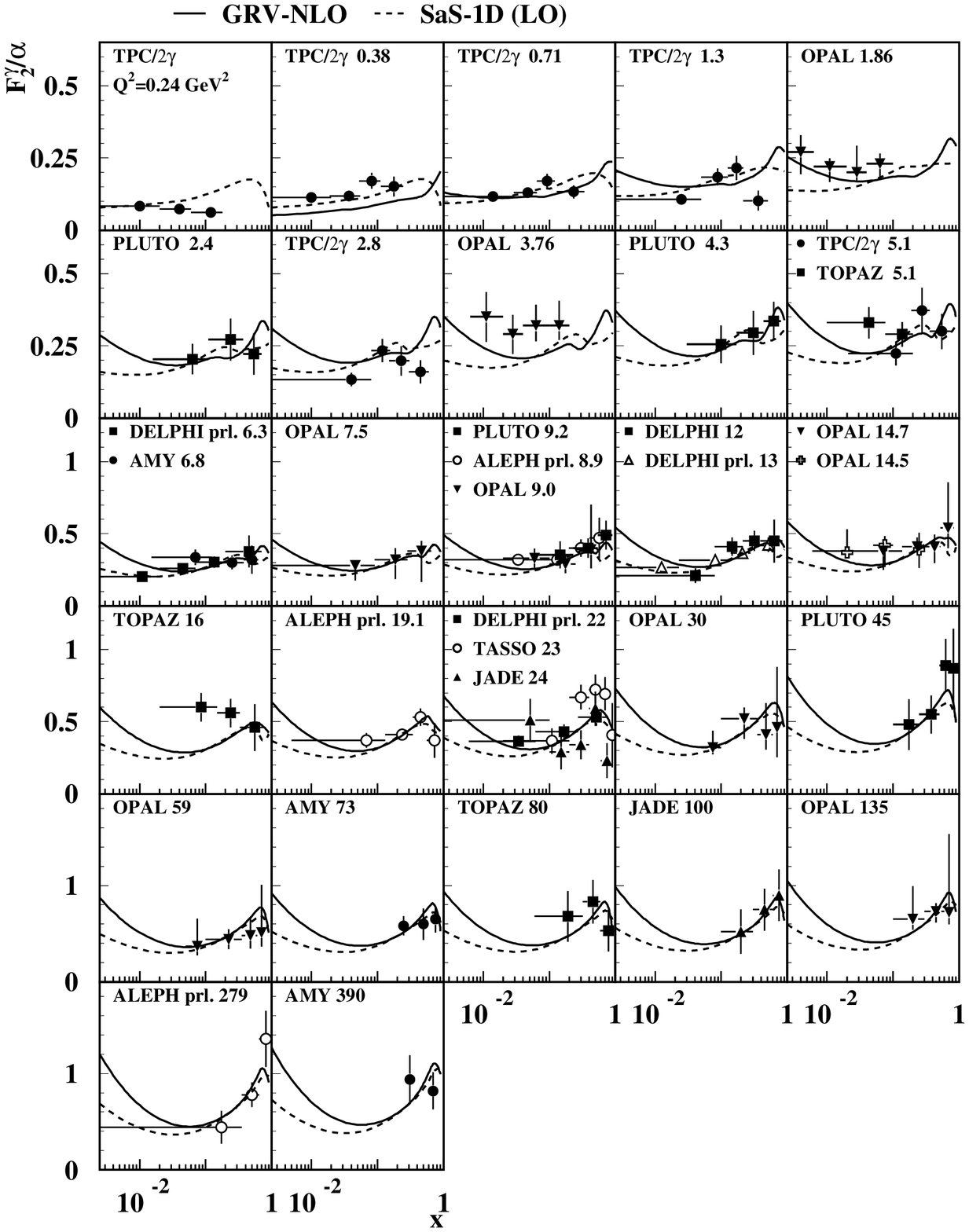,width=0.46\textwidth,height=7.1cm}
 &
\epsfig{file=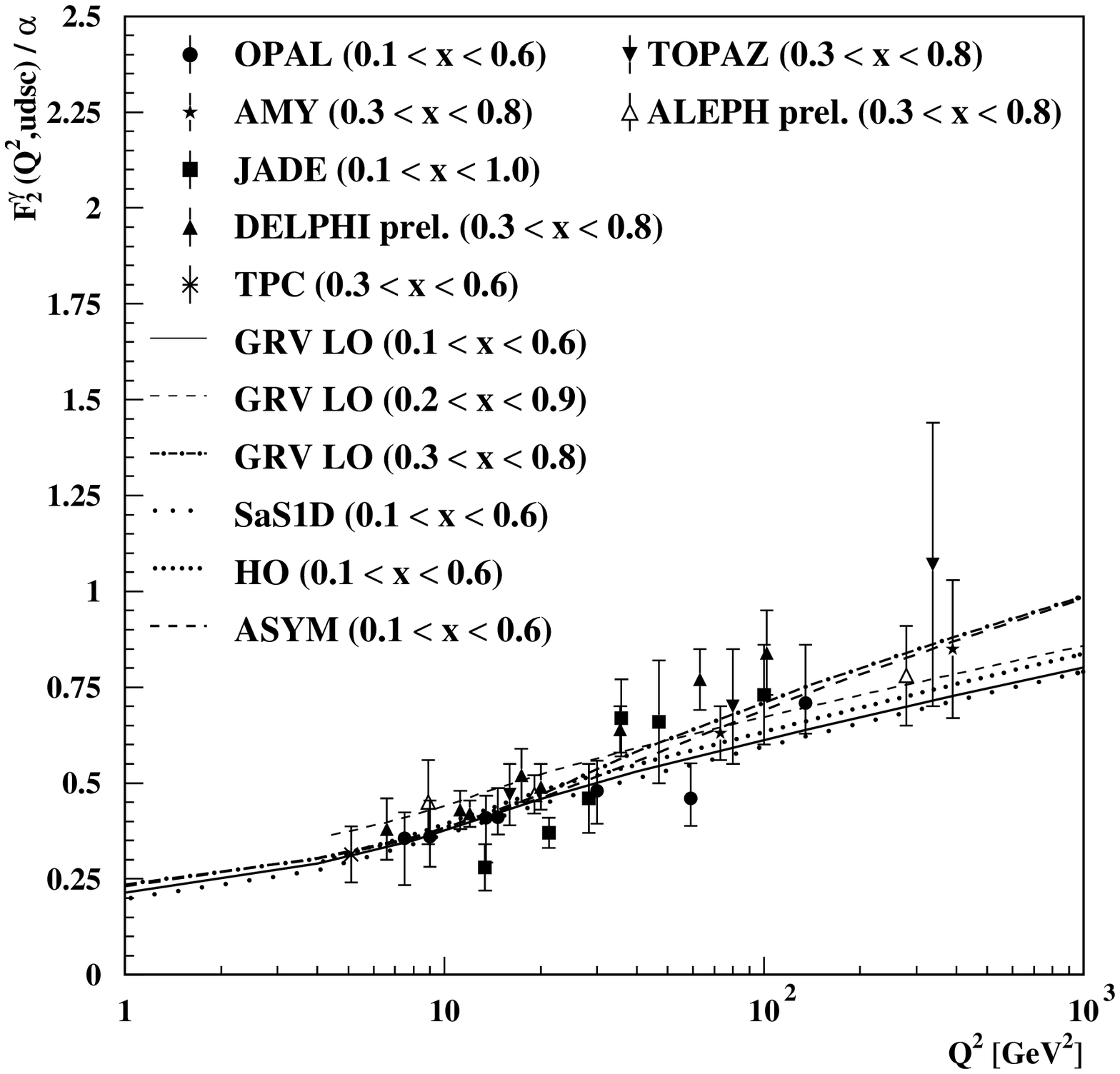,width=0.46\textwidth,height=7.1cm} 
\end{tabular}
\caption{\label{fig-sf}
Measurements of the photon structure function $F_2^{\gamma}$
in bins of $x$ and $Q^2$.
}
\end{figure}

The photon structure function $F_2^{\gamma}(x,Q^2)$ can be measured at LEP
in the range $x>10^{-3}$ and $1<Q^2<10^3$~GeV$^2$. This will make
it possible to
study the QCD evolution of $F_2^{\gamma}$ in a wide range
of $x$ and $Q^2$. All currently available measurements of the photon
structure function are shown in Fig.~\ref{fig-sf}~\cite{bib-f2g}. 
The data are compared to the next-to-leading order (NLO) GRV 
parametrisation \cite{bib-grv} and the leading order (LO) SaS-1D
parametrisation~\cite{bib-sas}. 

For low $x$ and not too small $Q^2$ a rise of the photon
structure function is expected, similar to the rise of the
proton structure function observed at HERA. 
An interesting new $F_2^{\gamma}$ measurement is presented by OPAL
in the $x$ and $Q^2$ ranges $2.5\times10^{-3}<x<0.2$ and $1.1<Q^2<6.6$~GeV$^2$.
The measurements is consistent with a possible rise within
large systematic errors. It should also be noted that
the OPAL points tend to be significantly higher than the previous
measurement by TPC/2$\gamma$ in a similar kinematic range.

For large $x$ and asymptotically large $Q^2$ the value of
$F_2^{\gamma}$ can be calculated from perturbative QCD due to the 
pointlike coupling of the photon to $\qqbar$ pairs~\cite{bib-witten}. 
The next-to-leading order (NLO) result~\cite{bib-buras} can 
be written as
\begin{equation}
\frac{F_2^{\gamma}}{\alpha}=\frac{a(x)}{\alpha_s(Q^2)}+b(x),
\end{equation}
where $a(x)$ and $b(x)$ are calculable functions which diverge
for $x\rightarrow 0$. The first term corresponds to the LO
result by Witten~\cite{bib-witten}. The measurement
of $F_2^{\gamma}$ could be a direct measurement of $\Lambda_{\rm QCD}$
if it were not for the large non-perturbative contributions
due to bound states. These contributions are large at all experimentally
accessible values of $Q^2$.

The evolution of $F_2^{\gamma}$ with $\ln Q^2$ is
shown in Fig.~\ref{fig-sf} using the currently available
$F_2^{\gamma}$ measurements for 4 active flavours. The data are compared to
the LO GRV and the SaS-1D parametrisations, and to a higher
order (HO) prediction based on the HO GRV parametrisation for
light quarks and on the NLO charm contribution calculated in
Ref.~13. The data are measured in different
$x$ ranges. The comparison of the LO GRV curves
for these $x$ ranges shows that for $Q^2>100$~GeV$^2$ significant
differences are expected.
An augmented asymptotic prediction for $F_2^{\gamma}$ is also
shown. The contribution to $F_2^{\gamma}$ from the three light flavours is 
approximated by Witten's leading order asymptotic form~\cite{bib-witten}.
This has been augmented by adding a charm contribution 
evaluated from the Bethe-Heitler formula~\cite{WIT-7601}, and an estimate of 
the hadronic part of $F_2^{\gamma}$, which essentially 
corresponds to the hadronic part of the GRV (LO) parametrisation.
In the region of medium $x$ values studied here this asymptotic prediction 
in general lies higher than the GRV and SaS predictions but it is still 
in agreement with the data.
The importance of the hadronic contribution to $F_2^{\gamma}$
decreases with increasing $x$ and $Q^2$, 
and it accounts for only 15~\% of $F_2^{\gamma}$ at 
$Q^2= 59$~GeV$^2$ and $x = 0.5$.

As predicted by QCD the evolution of $F_2^{\gamma}$ leads
to a logarithmic rise with $Q^2$, but theoretical and experimental
uncertainties are currently too large for a precision test of perturbative QCD.

\subsection{Azimuthal Correlations}
Only the structure function $F_2^{\gamma}$ has so far been determined
directly from measurements of double-differential cross-sections
for e$\gamma$ events with hadronic or leptonic final states.
It has been pointed out~\cite{bib-az} that azimuthal correlations 
in the final-state particles from two-photon collisions are
sensitive to additional structure functions.  
Azimuthal correlations can thus supplement the direct measurement of structure 
functions. ALEPH~\cite{bib-brew}, L3~\cite{bib-l3az} and OPAL~\cite{bib-azopal}
have measured azimuthal correlations using 
single-tag $\ee\rightarrow\ee\mu^+\mu^-$ events.

For single-tag events two independent angles can be defined
in the $\gamma\gamma^*$ centre-of-mass system 
assuming that the target photon direction is parallel
to the beam axis: The azimuthal angle $\chi$ is the angle between
the planes defined by the $\gamma\gamma^*$ axis and
the two-body final state and the $\gamma\gamma^*$ axis and the tagged electron.
The variable $\eta = \cos\theta^*$ is defined by the angle
$\theta^*$ between the $\mu^-$ and the $\gamma\gamma^*$ axis.

Neglecting the longitudinal component of the target photon
and setting $(1-y)$ to one, the cross-section can be written as
($F^{\gamma}_2 = 2xF^{\gamma}_T + F^{\gamma}_L$): 
\begin{equation}
\frac{{\rm d}\sigma ({\rm e\gamma \rightarrow e}\mu^+\mu^-)}
     {{\rm d}x {\rm d}y {\rm d}\eta {\rm d}\chi/2\pi} 
      =   \frac{2\pi\alpha^2}{Q^2} \left( \frac{1+(1-y)^2}{xy} \right) 
 \left[2x\tilde{F}^{\gamma}_T + 
                   \tilde{F}^{\gamma}_L -
                   \tilde{F}^{\gamma}_A \cos\chi +
                   \frac{\tilde{F}_B^{\gamma}}{2}\cos2\chi \right].  
\end{equation}
\begin{wrapfigure}[20]{r}{2.35in}
\epsfig{file=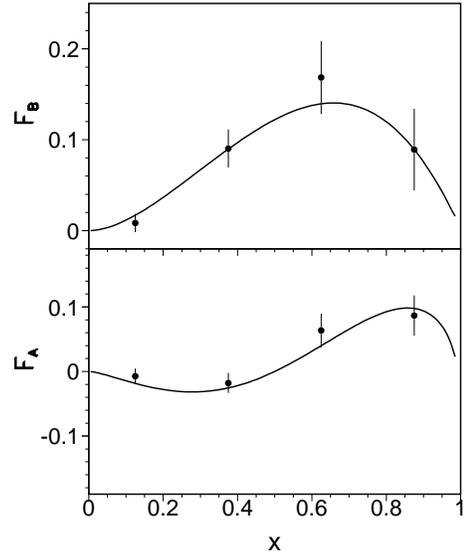,width=2.35in} 
\caption{\label{fig-l3fafb} The structure functions $F^{\gamma}_B$
and $F^{\gamma}_A$ for the process $\ee\rightarrow\ee\mu^+\mu^-$ (L3).
The lines show the QED expectation.}
\end{wrapfigure}
The conventional structure functions are 
recovered by integration over 
$\eta$ and $\chi$: $F_i^{\gamma}= \int_{-1}^{1} 
\int_0^{2\pi}\frac{d\chi d\eta}{2\pi} \tilde{F}_i^{\gamma} (i=2,A,B)$.
The structure functions $F_T$ and $F_L$ are related to the scattering of transverse and longitudinally polarized
virtual photons, respectively. $F_A$ is related to the
interference terms between longitudinal and transverse
photons and $F_B$ to the interference term between purely
transverse photons.
The longitudinal structure function \fl\ has been shown to be equal to the 
structure function \fb\ in leading order and for massless muons, 
although coming from different helicity states of the photons.  

\begin{figure}[htbp]
\begin{tabular}{cc}
\epsfig{file=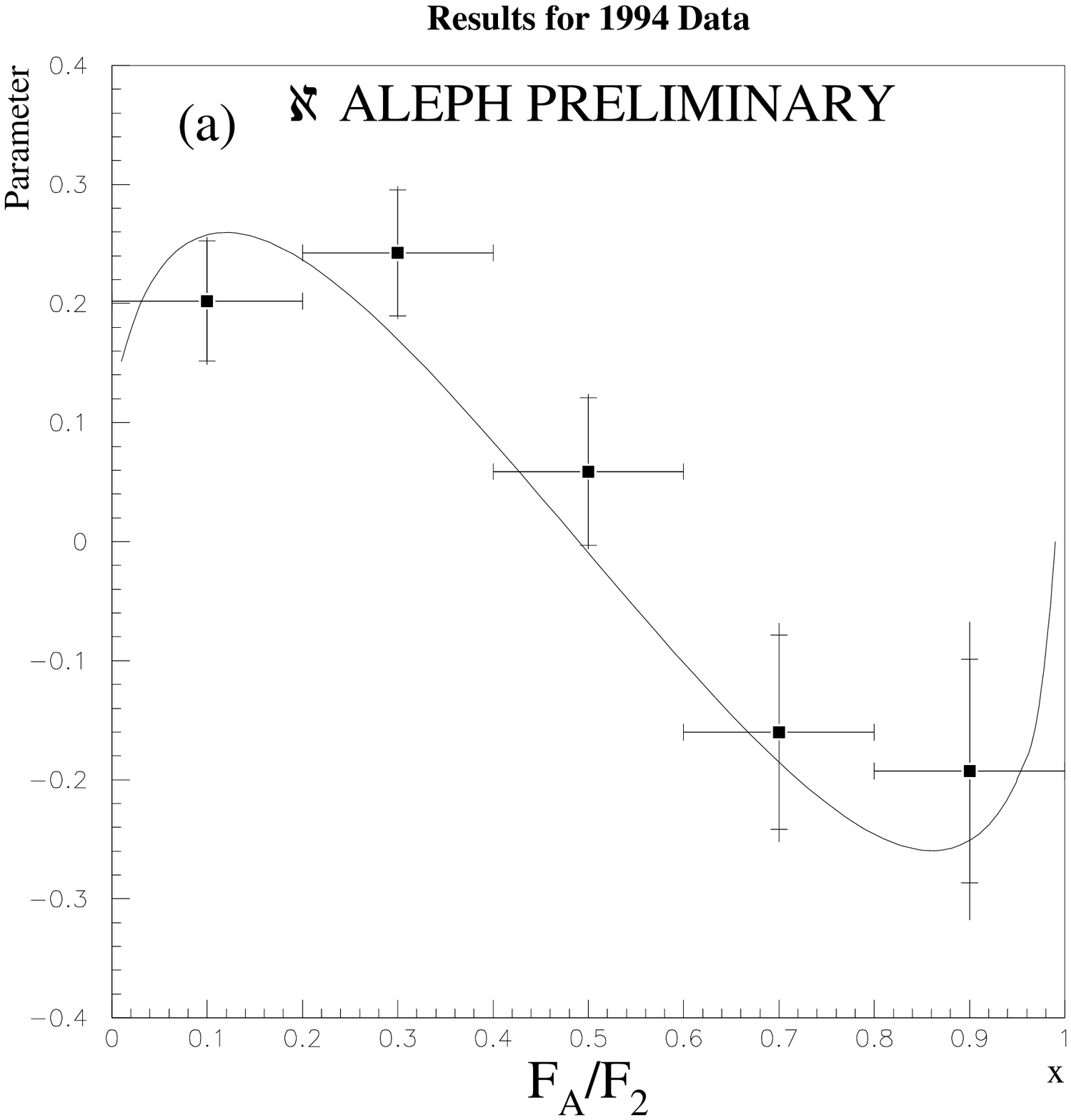,width=0.46\textwidth,height=6.5cm}
 &
\epsfig{file=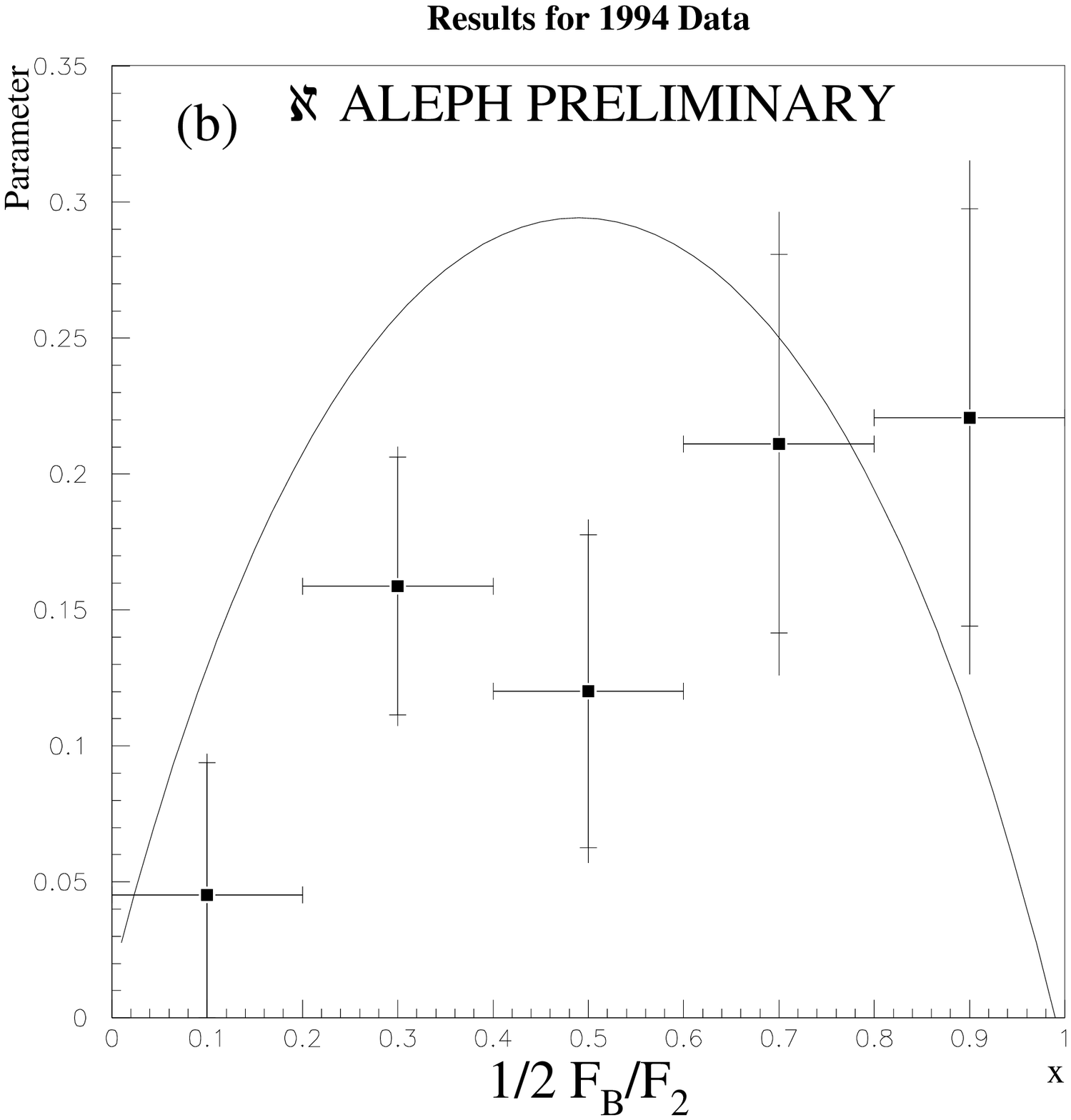,width=0.46\textwidth,height=6.5cm} 
\end{tabular}
\caption{\label{fig-brew} The ratio of structure functions, 
$F^{\gamma}_A/F_2^{\gamma}$ and $1/2\cdot F^{\gamma}_B/F_2^{\gamma}$, 
for the process $\ee\rightarrow\ee\mu^+\mu^-$ (ALEPH).
The lines show the QED expectation.}
\end{figure}

The variation of $F_A^{\gamma}$ and $F^{\gamma}_B$ with $x$ is 
in general consistent with QED (Figs.~\ref{fig-l3fafb} and \ref{fig-brew}
and Ref.~18).  
The measured values are significantly different from zero. 
Apart from being an interesting test of QED, these results
are especially important as a first step towards measuring
the additional structure function for hadronic events
using azimuthal correlations. Such a measurement will be
much more difficult due to the problems related to
the jet finding in hadronic events.

\section{Photon--Photon Scattering}
If both scattered electrons remain undetected, the $Q^2$
of the two interacting photons is very small and both
photons can be considered to be quasi-real. 
At high $\gg$ centre-of-mass energies $W=\sqrt{s}_{\gg}$ 
the total  cross-section for the production of hadrons in the interaction of 
two real photons is expected to be dominated by interactions
where the photon has fluctuated into an hadronic state. 

In LO QCD the $\gamma\gamma$ interactions can be classified
by three different processes: direct, single- and double-resolved. 
The bare photons interact in the direct process, whereas
in resolved events the partons (quarks or gluons) inside 
the hadronic state of the photon take part in the hard interaction. 
The probability to find partons in the photon is parametrised
by the parton density functions.

\subsection{Total Cross Sections}
Measuring the $\sqrt{s}_{\gg}$ dependence of the total $\gg$ cross-section
should improve our understanding of the hadronic nature of
the photon and the universal high energy behaviour of total 
hadronic cross-sections.

The total cross-sections $\sigma$ for hadron-hadron and $\gamma$p 
collisions are well described by a Regge parametrisation of the form
$\sigma=X s^{\epsilon}+Y s^{-\eta}$,
where $\sqrt{s}$ is the centre-of-mass energy of the hadron-hadron
or $\gamma$p interaction. The first term in the equation
is due to Pomeron exchange and the second term
is due to Reggeon exchange~\cite{bib-pdg}. Assuming factorisation of the 
Pomeron term $X$, the total $\gg$ cross-section can be related
to the pp (or $\ppbar$) and $\gamma$p total cross-sections at 
high centre-of-mass energies $\sqrt{s}_{\gg}$ where the Pomeron trajectory
should dominate:
\begin{equation}
\sigma_{\gg}
=\frac{\sigma_{\gamma{\rm p}}^2}{\sigma_{\rm pp }}.
\label{eq-tot2}
\end{equation}
A slow rise of the total cross-section with energy is predicted,
corresponding to $\epsilon\approx0.08$.
This rise can also be
attributed to an increasing cross-section for parton interactions
leading to mini-jets in the final state~\cite{bib-minijet}.

Before LEP the total hadronic $\gg$ cross-section has been measured
by PLUTO~\cite{bib-pluto}, TPC/2$\gamma$~\cite{bib-tpc} and
the MD1 experiment~\cite{bib-md1}. These experiments have measured
at centre-of-mass energies $W$ below 10~GeV before the onset 
of the high energy rise of the total cross-section.
Using LEP data taken at $\ee$ centre-of-mass energies $\sqee=130-161$~GeV
L3~\cite{bib-l3tot} has demonstrated that
the total hadronic $\gg$ cross-section in the range $5\le W \le 75$~GeV 
is consistent with the universal Regge behaviour of
\begin{wrapfigure}{l}{3in}
\epsfig{file=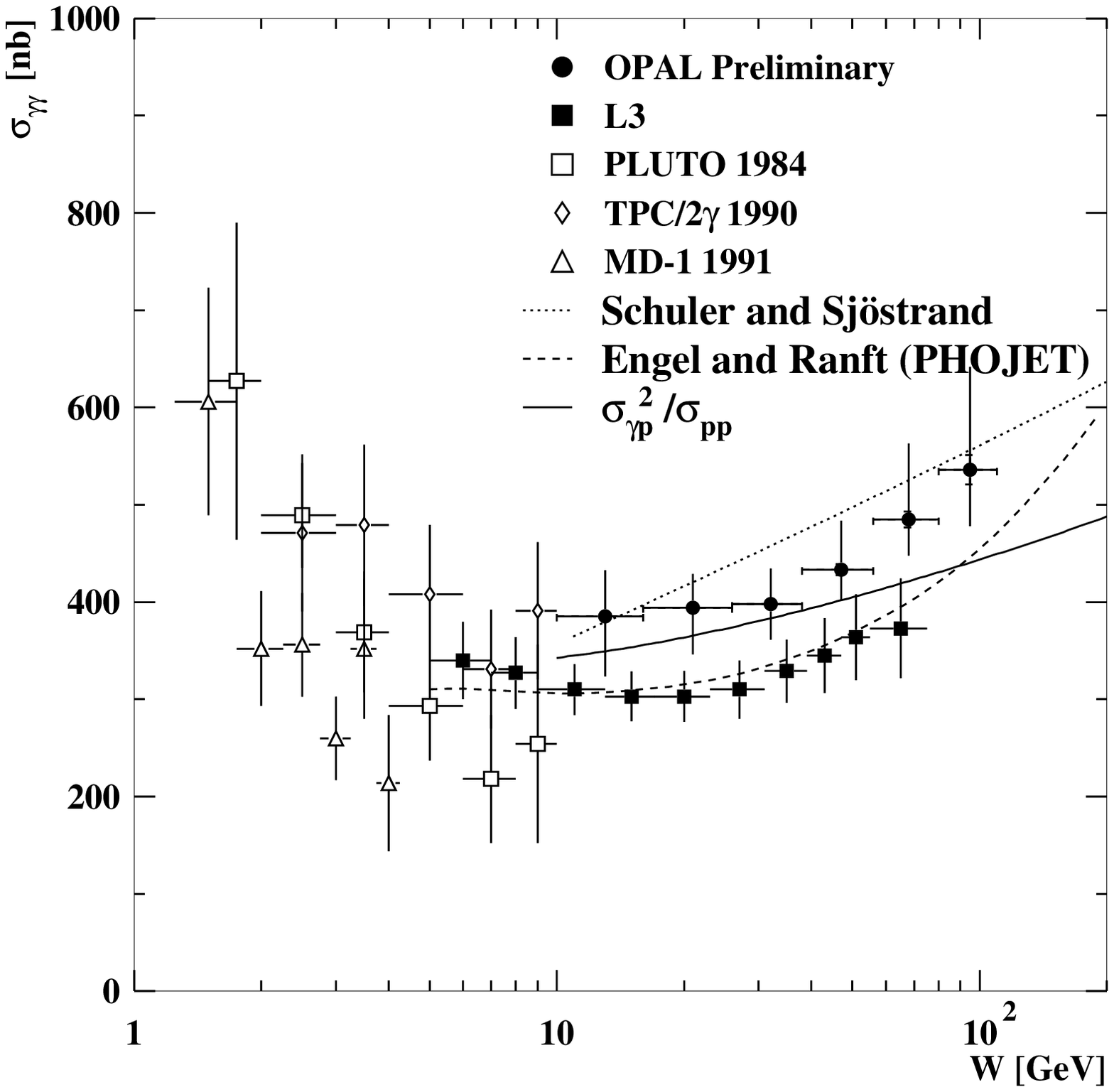,width=3in} 
\caption{\label{fig-stot} 
Total cross-section of the process $\gg\rightarrow\mbox{hadrons}$}
\end{wrapfigure}
total cross-sections which was also observed in $\gamma$p scattering at HERA.
The L3 measurement is shown in Fig.~\ref{fig-stot}
together with a preliminary OPAL measurement~\cite{bib-frank}
 in the range
$10<W<110$~GeV using data taken at $\sqee=161-172$~GeV.
The observed energy dependence of the cross-section is similar,
but the values for $\sigmagg$ are about 20~\% higher. 
It should be noted that the errors are strongly correlated between the $W$ bins
in both experiments. Furthermore, L3 has used the Monte Carlo
generator PHOJET~\cite{bib-phojet} for
the unfolding, whereas for the OPAL measurement 
the unfolding results of PHOJET and PYTHIA are
averaged. The unfolded cross-section using PHOJET
is about 5~\% lower than the central value.
In both experiments the cross-sections obtained
using PHOJET are lower than the cross-section obtained
with PYTHIA. 

The total cross-section is compared to several theoretical models.
Based on the Donnachie-Landshoff model~\cite{bib-DL}, 
the assumption of a universal high energy behaviour of 
$\gg$, $\gamma$p  and pp cross-sections is tested.
The parameters $X$ and $Y$ are fitted to the
total $\gg$, $\gamma$p and pp cross-sections 
in order to predict $\sigmagg$ via Eq.~\ref{eq-tot2}.
This is done assuming that the cross-sections can
be related at $\sqrt{s}_{\gg}=\sqrt{s}_{\rm \gamma p}=\sqrt{s}_{\rm pp }$.
The process dependent fit values for $X$ and $Y$ are
taken from Ref.~19 together with the values of the universal
parameters $\epsilon = 0.0790 \pm 0.0011$ and 
$\eta = 0.4678 \pm 0.0059$.
This simple ansatz gives a reasonable
description of the total $\gg$ cross-section $\sigmagg$.
Schuler and Sj\"ostrand \cite{bib-GSTSZP73} give a 
total cross-section for the sum of all possible event
classes in their model of $\gg$ scattering where the photon
has a direct, an anomalous and a VMD component.
They consider the spread between this prediction and
the simple factorisation ansatz as conservative estimate
of the theoretical band of uncertainty.
The prediction of Engel and Ranft~\cite{bib-phojet} is also plotted
which is implemented in PHOJET. It is in good agreement with
the L3 measurement and significantly lower than the OPAL
measurement. 
The steeper rise predicted by Engel and Ranft is in agreement
with both measurements.

A large part of the cross-section (about 20~\% in both MC models)
is due to diffractive and elastic events (e.g. $\gg\rightarrow\rho\rho$).
At high $W$ the detectors have only little acceptance
for these events and the correction procedure has to rely heavily on
the MC model. In the future it will therefore be very important
to gain a better understanding of these processes in $\gg$ interactions.

\subsection{Jet Production}
The measurement of inclusive jet cross-sections and the comparison
with NLO QCD calculations~\cite{bib-kleinwort,bib-aurenche} and LO QCD 
Monte Carlo simulations using different parametrisations of the 
parton distributions of the photons can constrain the relatively
unknown 
\begin{wrapfigure}[19]{r}{2.95in}
\epsfig{file=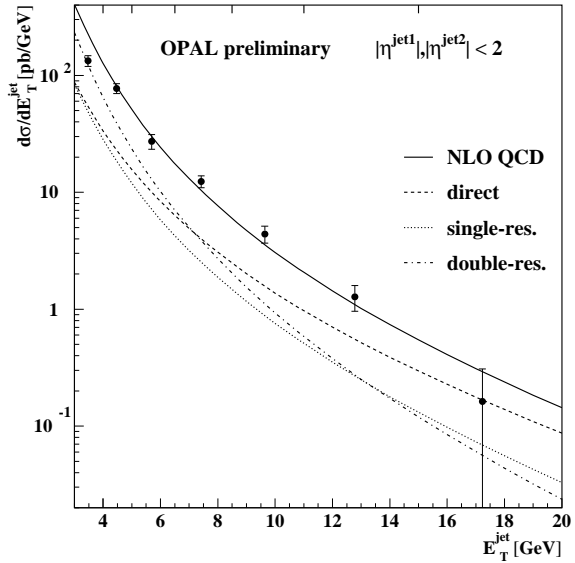,width=2.95in} 
\caption{\label{fig-ettwojet}The inclusive two-jet cross-section as a function
of $\ETJET$ for jets with $|\etajet|<2$ using a cone size $R=1$.}
\end{wrapfigure}
gluonic content of the photon.

In contrast to deep inelastic electron-photon scattering, which 
in leading order is only sensitive to
the quark content of the photon, the gluon content of the photon
can be tested, especially in the case of double-resolved processes,
in the interaction of two almost real photons.
Inclusive one-jet and two-jet cross-sections have been
measured at an $\ee$ centre-of-mass energy of $\sqee=58$ GeV at
TRISTAN~\cite{bib-amy,bib-topaz} and at $\sqee=130-172$ GeV
by OPAL~\cite{bib-opalgg,bib-opalgg2}. In all cases the cone
jet finding algorithm has been used.

In Fig.~\ref{fig-ettwojet} the $\ETJET$ distribution 
measured by OPAL~\cite{bib-opalgg2} at $\sqee=161-172$~GeV
is compared to a NLO perturbative QCD calculation of the 
inclusive two-jet cross-section
by Kleinwort and Kramer \cite{bib-kleinwort} who use
the NLO GRV parametrisation of the photon structure function \cite{bib-grv}.
The direct, single- and double-resolved parts and their sum are
shown separately. The data points are in good agreement with
the calculation except in the first bin where the calculation predicts 
a much higher cross-section. The resolved
cross-sections dominate in the region $\ETJET\;\sleq\;8$~GeV,
whereas, at high $\ETJET$ the direct cross-section is largest.
\begin{wrapfigure}{r}{3in}
\epsfig{file=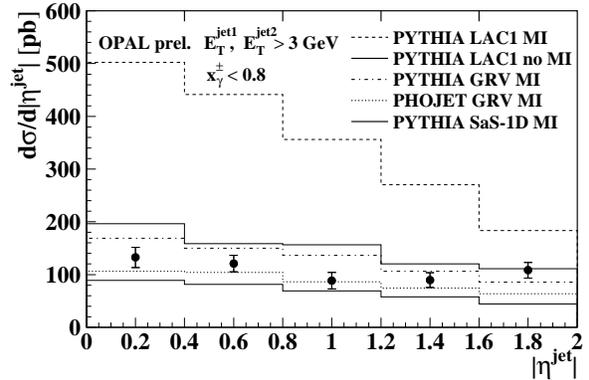,width=3in} 
\caption{\label{fig-etatwojet}The inclusive two-jet cross-section as a function
of $\etajet$ for jets in mainly resolved event
with $\ETJET>3$~GeV using a cone size $R=1$.}
\end{wrapfigure}

The NLO QCD calculation gives the jet cross-section
for massless partons, whereas the 
experimental jet cross-sections are measured for hadrons. 
The difference due to fragmentation mainly contributes at low $\ETJET$ values.
A correction to the parton level would increase the 
cross-section by about a factor of 1.2 to 1.3.

The inclusive two-jet cross-section as a function of $|\etajet|$
is shown in Fig.~\ref{fig-etatwojet} for events with a large double-resolved 
contribution after requiring $\xgpm<0.8$. The variable
$\xgpm$ specifies the fraction of the photon energy participating in the
hard scattering:
\begin{equation}
\xgp=\frac{\displaystyle{\sum_{\rm jets=1,2}(E+p_z)}}
 {{\displaystyle\sum_{\rm hadrons}(E+p_z)}} \;\;\;\mbox{and}\;\;\;
\xgm=\frac{\displaystyle{\sum_{\rm jets=1,2}(E-p_z)}}
{\displaystyle{\sum_{\rm hadrons}(E-p_z)}},
\label{eq-xgpm}
\end{equation}
where $p_z$ is the momentum component along the $z$ axis of the
detector and $E$ is the energy of the jets or hadrons.
Ideally, for direct events without remnant jets $\xgp=1$ and $\xgm=1$,
whereas
for double-resolved events both values $\xgp$ and $\xgm$
are expected to be much smaller than~1.

The inclusive two-jet cross-section predicted by the two
LO QCD Monte Carlo models, PYTHIA and 
PHOJET differ significantly even if the same 
photon structure function (here GRV LO) is used. 
This model dependence reduces the
sensitivity to the parametrisation of the photon structure function.
Different parametrisations were used as input to the PYTHIA simulation. 
The GRV-LO~\cite{bib-grv} and SaS-1D parametrisations~\cite{bib-sas} describe
the data equally well, but LAC1~\cite{bib-LAC1} which increases
the cross-section for gluon-initiated processes
overestimates the inclusive two-jet cross-section
significantly. Turning off the simulation of multiple interactions (MI)
within PYTHIA reduces the predicted cross-section using LAC1 by
more than a factor of two.

For a more quantitative interpretation of these results
in terms of parton distribution functions, it
will be very important to understand the influence of
multiple interactions on the jet cross-sections and
to use jet definitions which will allow to compare
theory (partons) and experiment (hadrons) directly. This
is very similar to the problems of measuring jet cross-sections in 
photoproduction at HERA discussed in these 
proceedings~\cite{bib-butter,bib-klasen}. 

\subsection{Charm Production}
Similar to jet production, open charm production in photon-photon collisions 
can also be used to constrain the gluon content of the photon.
The charm production cross-sections have been calculated
in NLO~\cite{bib-drees,bib-cacc}. 
The mass $m_{\rm c}$ of the charm quark sets
the scale for the perturbative QCD calculation. The cross-section
is factorized into the matrix elements for the production
of heavy quarks and the parton densities for light quarks (q) and
gluons. This `massive' approach is expected to be valid
if the transverse momenta $p_{\rm T}$ of the charm quarks are of the
same order, $p_{\rm T} \approx m_{\rm c}$, which is true
for the kinematic range probed at LEP.
At LEP energies only the direct process $\gg\rightarrow\ccbar$ and the
single-resolved process $g$q$\rightarrow\ccbar$ are relevant.

The cleanest method to tag open charm is the reconstruction of 
$D^{*+}\rightarrow D^0\pi^+$ decays. Due to the small branching ratios
of the $D^0$ into charged pions and kaons, this 
\begin{wrapfigure}{r}{3in}
\epsfig{file=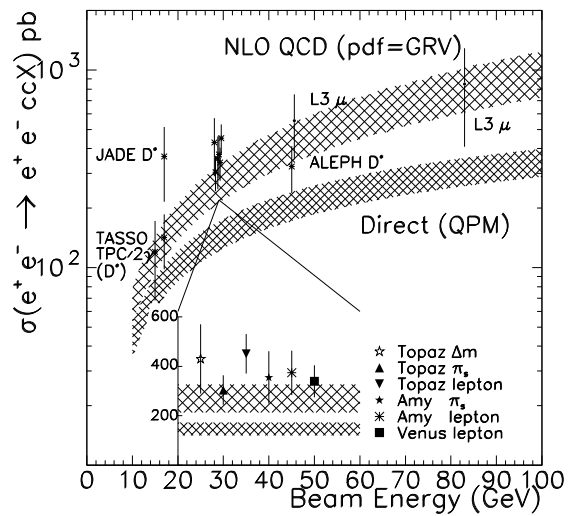,width=3in}
\caption{\label{fig-charm}
Cross-section for the process $\ee\rightarrow\ee\mbox{c}
\overline{\mbox{c}}$ as a function of the electron beam energy.} 
\end{wrapfigure}
method is statistics limited. ALEPH has measured the charm cross-section
using $33\pm8$ $D^{*\pm}$ mesons reconstructed in the LEP1 data.
L3 has measured the charm cross-section in $\gg$ interaction
at LEP1 and LEP2 by tagging muons from semi-leptonic charm decays
in the momentum range $2<p_{\mu}<7$~GeV$/c$ at LEP1 and
$2<p_{\mu}<15$~GeV/$c$ at LEP2~\cite{bib-l3charm}. The efficiency to tag muons
is less than $10^{-3}$ leading to large systematic and
statistical uncertainties. 

The cross-section for the process $\ee\rightarrow\ee\mbox{c}
\overline{\mbox{c}}$ as a function of the beam energy is shown
in Fig.~\ref{fig-charm}. The experimental results for
various charm tagging methods used by pre-LEP
experiments have been extrapolated to obtain a total charm
cross-section~\cite{bib-lep2}. 
The upper band shows the full NLO charm cross-section
calculated by Drees et al~\cite{bib-drees} and
the lower band the contribution from the Born term direct process
(Quark Parton Model, QPM). The upper edge of the band
is obtained by setting $m_{\rm c}=1.3$~GeV with a  scale $\mu=m_{\rm c}$ 
and the lower edge by setting $m_{\rm c}=1.7$~GeV with $\mu=\sqrt{2}
m_{\rm c}$.
The data points obtained from the TRISTAN and JADE measurements cluster
around the higher edge of the the `massive' NLO calculation 
which uses the GRV parametrisation. At LEP energies
the extrapolated measurement are in good agreement with 
the NLO calculation within the large errors.

\section{Conclusions}
Measurements of the photon structure function $F_2^{\gamma}(x,Q^2)$
from double-differential cross-section in e$\gamma$ scattering mainly 
constrain the quark distribution in the photon. 
At low $x$ LEP will be able to study the region where the onset of the rise of
$F_2^{\gamma}$ is expected from the HERA data on the
proton structure function. 
LEP measurements will cover the kinematic range $x>10^{-3}$ and 
$1<Q^2<10^3$~GeV$^2$. 
A first measurement in the low $x$ region has been presented by OPAL.

The logarithmic rise of $F_2^{\gamma}$ with $Q^2$ for
medium $x$ and large $Q^2$ is observed as predicted, 
but theoretical and experimental
uncertainties are currently too large for a precision test of perturbative QCD.
The current systematic limitation of the $F_2^{\gamma}$ measurements
comes from the discrepancies between data distributions and Monte Carlo models
for the hadronic final state. Work has started
to improve the Monte Carlo models.

Azimuthal correlations have been used to determine
the structure functions $F_A$ and $F_B$ in single-tagged
$\ee\rightarrow\ee\mu^+\mu^-$ events. These structure
functions correspond to different helicity states of
the virtual photon and the target photon. The results
are consistent with QED. In the future it will be interesting
to extend these studies to hadronic final states.

First measurements by L3 and OPAL of the energy dependence of the total
cross-section for hadron production in the interactions
of quasi-real photons show the slow rise characteristic
for hadronic interactions. However, the cross-section measured
by L3 is about 20~\% lower than the cross-section measured
by OPAL. 

Measurements of jet production in $\gg$ interactions
presented by OPAL disfavour parametrisations with
a large gluon content in the photon like LAC1. The cross-sections
are in good agreement with NLO calculations using the GRV
parametrisation. Good agreement with the NLO calculation is also found
for the measurement of the total charm production cross-section
in $\gg$ interactions by ALEPH and L3.

The experimental methods and the theoretical framework necessary
for studying photon structure in the interactions of quasi-real
photons at LEP and in photoproduction at HERA are very similar.
This workshop has shown very clearly that both communities
can learn a lot from each other.

\section{Acknowledgements}
I want to thank the Max-Planck-Institut f\"ur Physik
and DESY, especially Bernd Kniehl, Gustav Kramer and Albrecht Wagner, 
for the excellent organisation
of this interesting workshop. Thanks also to the LEP collaborations
for providing the plots.

\end{document}